\documentclass[a4paper,12pt]{article}

\usepackage[a4paper,top=2cm,bottom=2cm,left=3cm,right=2cm]{geometry}
\usepackage{amssymb,graphicx}
\usepackage{epstopdf}
\usepackage{amsmath,amsfonts}
\usepackage{mathtools}
\usepackage{hyperref}
\usepackage[dvipsnames]{xcolor}
\usepackage{caption}
\usepackage{subcaption}
\RequirePackage[numbers,sort&compress]{natbib}

\usepackage[T1]{fontenc}

\def\d{\partial}

\newcommand{\be}{\begin{equation}}
\newcommand{\ee}{\end{equation}}
\newcommand{\bea}{\begin{eqnarray}}
\newcommand{\eea}{\end{eqnarray}}
\newcommand{\bg}{\begin{gather}}
\newcommand{\eg}{\end{gather}}
\newcommand{\bseq}{\begin{subequations}}
\newcommand{\eseq}{\end{subequations}}

\def\half{\frac{1}{2}}

\newcommand{\pd}[2]{\ensuremath{\frac{\d #1}{\d #2}} }

\DeclareMathOperator{\tr}{tr}

\newcommand{\pt}{\partial}

\newcommand{\I}{\mathcal I}

\numberwithin{equation}{section}

\begin{document}

\vspace{.5cm}

\begin{center}
{\Large\sc Viscosity bound violation in holographic solids \\[5mm]
and the viscoelastic response}

\vspace{1.cm}

\textbf{Lasma Alberte$^a$, Matteo Baggioli$^{b,c}$, Oriol Pujol{\`a}s$^b$
}\\

\vspace{1.cm}
${}^a\!\!$ {\em {Abdus Salam International Centre for Theoretical Physics (ICTP)\\Strada Costiera 11, 34151, Trieste, Italy}}

\vspace{.2cm}
${}^b\!\!$ {\em {Institut de F\'isica d'Altes Energies (IFAE)\\ 
The Barcelona Institute of Science and Technology (BIST)\\
Campus UAB, 08193 Bellaterra (Barcelona) Spain
}}

\vspace{.2cm}
${}^c\!\!$ {\em {
Department of Physics, Institute for Condensed Matter Theory\\
University of Illinois, 1110 W. Green Street, Urbana, IL 61801, USA
}}

\end{center}

\vspace{0.8cm}

\centerline{\bf Abstract}
\vspace{2 mm}
\begin{quote}\small

We argue that the Kovtun--Son--Starinets (KSS) lower bound on the viscosity to entropy density ratio holds in fluid systems but is violated in solid materials with a non-zero shear elastic modulus. 
We construct explicit examples of this by applying the standard gauge/gravity duality methods to massive gravity and show that the KSS bound is clearly violated in black brane solutions whenever the massive gravity theories are of {\em solid} type. 
We argue that the physical reason for the bound violation relies on the viscoelastic nature of the mechanical response in these materials.
We speculate on whether any real-world materials 
can violate the bound and discuss a possible generalization of the bound that involves the ratio of the shear elastic modulus to the pressure.

 \end{quote}

\vfill
 
\newpage

\tableofcontents

\section{Introduction}

It has been long known that black 
brane solutions can be characterized both by thermodynamic quantities like temperature and entropy as well as hydrodynamic entities like viscosity and diffusion. In gauge/gravity duality \cite{Maldacena:1997re}, the hydrodynamics of the black branes is mapped to the hydrodynamic properties in the dual field theory. One of the most prominent insights that the AdS/CFT correspondence have provided for the understanding of dynamics of strongly coupled condensed matter systems is that the shear viscosity to entropy density ratio takes on a universal value for all gauge theories with Einstein gravity duals \cite{Policastro:2001yc,Kovtun:2004de}:\footnote{We work in the units where $\hbar=k_B=8\pi G\equiv1$.} 
\be\label{univ}
\frac{\eta}{s}=\frac{1}{4\pi}\;.
\ee
This value was conjectured to set a fundamental lower bound on this ratio --- the celebrated Kovtun--Son--Starinets (KSS) bound \cite{Kovtun:2004de}. Besides the universality of the viscosity to entropy ratio in all Einstein gravity duals \cite{Kovtun:2003wp,Iqbal:2008by} there are several other arguments supporting this claim. 
Above all, the bound seems to be satisfied for all known fluids where $\eta/s$ has been measured, including examples like superfluid helium \cite{Rupak:2007vp} and the QCD quark gluon plasma  (see \emph{e.g.} \cite{Song:2010mg}). 

By now it is well established that the KSS bound is violated by higher curvature corrections to the Einstein theory. In particular, the violation of the bound was observed in Einstein gravity supplemented by the quadratic Gauss--Bonnet term \cite{Brigante:2007nu}. In terms of the Gauss--Bonnet coupling $\lambda_{GB}$ the viscosity to entropy density ratio was found to be
\be
\frac{\eta}{s}=\frac{1}{4\pi}\left[1-4\lambda_{GB}\right]\;.
\ee
For a positive coupling this would imply an arbitrary violation of the bound. However, the consistency requirements on the dual field theory impose constraints on the allowed values of the Gauss--Bonnet coupling constant. In particular, it was found that the field excitations in the dual field theory allow for superluminal propagation velocities for $\lambda_{GB}>9/100$, thus imposing a new lower bound on the viscosity to entropy ratio \cite{Brigante:2008gz}. In the light of these results it is at present not clear whether a universal fundamental bound on the shear viscosity to entropy ratio exists. For a review on the bound violation in higher derivative theories of gravity, see \cite{Cremonini:2011iq} and references therein. Another example of violation of the bound has been found in anisotropic theories \cite{Rebhan:2011vd,Jain:2014vka,Critelli:2014kra,Ge:2014aza}.

In this work we shall study the possibility of violating the KSS bound in the field theories dual to massive gravitational theories. In distinction from Gauss--Bonnet gravity, introducing a non-vanishing graviton mass modifies gravity in the infrared and therefore this can have a large impact on the homogeneous and static response.  In holographic context, massive gravity has already turned out to provide a useful mechanism of incorporating momentum dissipation \cite{Vegh:2013sk,Blake:2013bqa}. It was also shown that although the energy-momentum tensor is not conserved in field theories dual to massive gravity, it nevertheless admits an effective hydrodynamic description at sufficiently high temperatures and small values of graviton mass \cite{Davison:2013jba,Davison:2014lua}. Here we shall use the most general massive gravity models that can be written in terms of two St\"uckelberg scalar fields first introduced in \cite{Baggioli:2014roa} (for earlier related works, see \cite{Andrade:2013gsa,Taylor:2014tka}). Basing on the analogy to the flat space effective field theory (EFT) description of solids and fluids \cite{Leutwyler:1993gf,Leutwyler:1996er,Dubovsky:2011sj,Nicolis:2015sra,Nicolis:2013lma} it has been argued recently that these two fields models of massive gravity can also be broadly divided into solids and fluids \cite{Alberte:2015isw}. The claim was further supported by studying the response of the massive gravity black branes to external shear strain deformations: massive gravities with the symmetries of the solid EFTs are endowed with a non-zero shear elastic modulus. The fluid massive gravities, however, exhibit zero elastic response. A precise definition of solid and fluid massive gravities will be given in section \ref{sec:solids}.

The purpose of this work is to analyze the impact of the graviton mass term on the viscosity and on the KSS bound in the case of solid massive gravity black branes. These are holographically dual to solids that we shall refer to as \emph{holographic solids} henceforth.
Naively, one might expect that solids correspond to the large viscosity limit, $\eta\to\infty$, and therefore does not exhibit any reduction in the ${\eta}/{s}$ ratio. However, this is not the case. 
The characteristic response under mechanical deformations of solids is very different from that of fluids, especially under static and homogeneous deformations. A material that exhibits an elastic response, \emph{i.e.} a solid, counteracts an applied constant stress with a constant in time deformation characterized by a displacement vector, $u_i$, and a constant strain tensor, $\partial_{(i}u_{j)}$. In a viscous fluid, in turn, a constant applied stress results in a constant flow --- a constant velocity gradient or strain rate, $\partial_{(i}\dot u_{j)}$. The response under mechanical deformations can actually be more complex and exhibit both types of behaviour. This is what happens in \emph{viscoelastic materials}, a more precise definition of which we defer for the next section. Heuristically, it is clear that if certain massive gravity black branes are endowed with a notion of elasticity then they must be viscoelastic at least in the limit of small graviton mass where one recovers the Einstein gravity with viscosity given by \eqref{univ}. Below we investigate how turning on a finite elastic response affects the viscosity and show that the KSS bound is violated in theories dual to solid massive gravities. 

\section{Viscoelasticity}\label{sec:viscous}

In the standard mechanical linear response theory \cite{landau7,Lubensky} the internal stress of a homogeneous and isotropic material due to a constant shear deformation described by the displacement vector $u_i$ can be expressed via the linear relation
\be
T_{ij}^{(T)}=G\,u_{ij}^{(T)}\;
\ee
between the traceless stress tensor and the traceless part of the linear displacement tensor
\be
u_{ij}=\frac{1}{2}\left(\pt_i u_j+\pt_j u_i\right)\;.
\ee
The proportionality coefficient $G$ in the shear stress/strain ratio is the modulus of rigidity and is non-zero only for solid materials, which are said to exhibit an {\em elastic} response. 

Fluid materials, instead, show a viscous response whereby the constant internal stress is due to a deformation with a constant shear rate,
\be
T_{ij}^{(T)}=\eta\,\dot u_{ij}^{(T)}\;.
\ee
The shear stress/strain rate ratio defines the viscosity of the material. 


Materials that exhibit a viscoelastic response display both elastic and viscous fluid behaviour. The simplest way to introduce such materials is by considering a general time dependence in $u_{ij}$ (and in $T_{ij}^{(T)}$). The most common presentation of viscoelasticity then consists of a low frequency response relation of the  form \cite{landau7}
\be\label{viscoel}
T_{ij}^{(T)}=G\,u_{ij}^{(T)}+\eta \,\dot u_{ij}^{(T)}\;.
\ee
Some of the typical viscoelastic phenomena  such as {\em creep} and {\em stress relaxation} follow easily from \eqref{viscoel}: the system relaxes to an equilibrium configuration 
within a `Maxwell' relaxation time given by
$$
\tau = {\eta \over G}~.
$$
It is clear that \eqref{viscoel} can be understood as a truncated expansion in time derivaties of a kind of damped oscillator, and that the  most important parameters that 
encode the viscoelastic response at low frequencies are the modulus of rigidity, $G$,  and the shear viscosity, $\eta$.

For both the modulus of rigidity and the shear viscosity, one can define convenient notions of these transport coefficients from Green--Kubo (GK) formulas that relate them to 
the correlators of the stress tensor \cite{landau9}.
The GK shear viscosity is thus given by \cite{Policastro:2001yc,Kovtun:2004de}
\be\label{viscosity}
\eta \equiv \lim_{\omega\to 0}\frac{1}{\omega}\text{Im }{\cal G}^R_{T_{ij}\,T_{ij}}\;
\ee
where ${\cal G}^R$ is the retarded Green's function of the stress tensor. The GK modulus of rigidity can similarly be defined  as \cite{Alberte:2015isw}
\be\label{elasticity}
G \equiv \lim_{\omega\to 0}\text{Re }{\cal G}^R_{T_{ij}\,T_{ij}}\;.
\ee
In the rest of this work we shall refer only to these GK notions of viscosity and rigidity, which are straightforward to obtain using the standard gauge/gravity prescription for the holographic retarded correlators.

In terms of the two parameters defined in \eqref{viscosity} and \eqref{elasticity}, the static mechanical response of generic isotropic materials can be depicted in the $\{G,\eta\}$ plane. The $G=0$ axis corresponds to fluids. The $\eta=0$ axis to non-dissipative ({\emph{e.g.} at zero temperature)  solids. The rest of the two dimensional space is spanned by viscoelastic materials. As we shall see, solids dual to massive gravity black branes do lie inside this plane.

\section{Holographic Solids}\label{sec:solids}
We consider a $3+1$ dimensional gravity theory
\be\label{action}
S=\int d^4x\,\sqrt{-g}\left[\frac{1}{2}\left(R+\frac{6}{L^2}\right)-\frac{L^2}{4}F_{\mu\nu}F^{\mu\nu}-\frac{m^2}{L^2}\,V(X,Z)\right]+\int_{r\to0}d^3x\sqrt{-\gamma}\,K\;,
\ee
where $L$ is the AdS radius, $m$ is a dimensionless mass parameter, and
\begin{align}
X  \equiv \half \tr[\I^{IJ}]\,,\qquad  Z \equiv \det [\I^{IJ}]\,,\qquad \mathcal I^{IJ}\equiv\d_\mu \phi^I \d^\mu \phi^J\,, 
\end{align}
and the indices $I,J=\{x,y\}$ are contracted with $\delta_{IJ}$. In \eqref{action}, we have included the Gibbons--Hawking boundary term where $\gamma$ is the induced metric on the AdS boundary, and $K=\gamma^{\mu\nu}\nabla_\mu n_\nu$ is the extrinsic curvature with $n^\mu$ --- an outward pointing unit normal vector to the boundary. Around the scalar fields background $\hat\phi^I=\delta^I_ix^i$ the metric admits the black brane background solution 
\be
\label{solmetric}
ds^2 =   L^2\left(\frac{dr^2}{f(r)r^2}+\frac{-f(r)dt^2+dx^2+dy^2}{r^2}\right) \;,
\ee
with the emblackening factor given in terms of the background value of the mass potential: 
\begin{equation}\label{f2phi}
f(r) = 1 +  \frac{\mu^2 r^4}{2 r_h^2}+m^2\,r^3 \, \int^r d\tilde r\frac{1}{\tilde r^4} \hat V(\tilde r)\; ,
\end{equation} 
where $\hat V(r) \equiv V(\hat X, \hat Z)$. The solution for the Maxwell field is $\hat A_t = \mu\left(1-r/r_h\right)$.

The viscoelastic response of the boundary theory in the holographic description is encoded in the transverse traceless tensor mode of the metric perturbations.\footnote{We define the metric perturbations as $g_{\mu\nu}=\hat g_{\mu\nu}+h_{\mu\nu}$ so that the inverse perturbations $g^{\mu\nu}=\hat g^{\mu\nu}+h^{\mu\nu}$ are given by $h^{\mu\nu}=-\hat g^{\mu\alpha}\hat g^{\nu\beta}h_{\alpha\beta}+\mathcal O(h^2)$.} In $3+1$ dimensional bulk, tensor modes exist only for perturbations homogenous in the transverse directions and its two helicity-two components can be parametrised as $h_+ \equiv \half \frac{L^2}{r^2} \big(h^{xx} - h^{yy} \big)$ and $h_\times \equiv  \frac{L^2}{r^2} h^{xy}$. The quadratic action for the homogeneous tensor mode takes the form
\begin{equation}\label{Stensor}
S = \int d^4 x\, \frac1{4}\frac{L^2}{r^2} \left( \frac1{f(r)} (\dot h)^2 - f(r) (h')^2 - 4m^2M^2(r) \frac{r^2}{L^2}\, h^2\right) \; ,
\end{equation}
where $h$ stands for any of the two components $h_+, h_\times$ and we have defined a mass function 
\begin{equation} \label{m_ten}
M^2(r) \equiv \frac1{2r^2} \hat V_X(r) \;.
\end{equation}
The subscript $X$ denotes the partial derivative $V_X \equiv \pd V X$. The equation of motion for the Fourier mode, defined through $h(t,r)=\int\frac{d\omega}{2\pi}\,h_\omega(r)e^{-i\omega t}$, becomes that of a massive scalar field living on the black brane background \eqref{solmetric}:
\be\label{spin2}
\left[f\partial_r^2 +\left(f'-2\frac{f}{r}\right)\partial_r + \left(\frac{\omega^2}{f} -4m^2 M^2(r)\frac{r^2}{L^2}\right)\right]h_\omega=0 \; .
\ee
It is very important to emphasize here that the mass of the tensor mode is only due to the $X$ dependence of the potential $V(X,Z)$. Hence, in the case when $V$ is only a function of $Z$ the graviton remains massless. In our previous work we have argued that in the case when $V=V(Z)$ the dual theory describes \emph{fluids}, whereas the presence of an $X$ dependence, \emph{i.e.} when $V=V(X,Z)$, indicates that the material is a \emph{solid} \cite{Alberte:2015isw}. We have also shown that there is no elastic response in the case of fluids. Moreover, since for fluids the graviton mass is zero, the universality proof \cite{Iqbal:2008by}  for the viscosity to entropy ratio based on the membrane paradigm is applicable and we expect no violation of the KSS bound. Without loss of generality we therefore only consider the theories describing solids with graviton mass terms of the form
\be\label{vx}
V(X)=X^n\;.
\ee
Here we are allowing for general values of $n$ in order to see what is the impact of this parameter on the elasticity and viscosity. There is a naturally preferred value, $n=1$, where the model enjoys enhanced consistency properties in the sense that the St\"uckelberg sector does not introduce a low strong coupling scale (see \cite{Alberte:2014bua,Baggioli:2014roa,Alberte:2015isw} for more consistency constraints).

On the background solution $\hat X=(r/L)^2$, and the graviton mass potential \eqref{vx} yields a mass function
\be\label{mass0}
M^2(r)=\frac{n}{2L^2}\left(\frac{r}{L}\right)^{2(n-2)}=\frac{n}{2L^2}\left(\frac{r}{L}\right)^\nu\;,\qquad\text{so that}\qquad n=\frac{4+\nu}{2}\;.
\ee
The emblackening factor in this case becomes 
\be\label{fsol}
f(r)=1+\frac{\mu^2r^4}{2r_h^2}+\frac{m^2}{\nu+1}\left(\frac{r}{L}\right)^{\nu+4}-Mr^3\;.
\ee
We note that this expression is not valid for the case $\nu=-1$. The integration constant $M$ is determined from the condition $f(r_h)=0$ and up to an additional constant is proportional to the energy density $\varepsilon$ in the renormalized dual field theory, \emph{i.e.} $\varepsilon=ML^2$, and reads:
\be\label{endens}
\varepsilon=\frac{L^2}{r_h^3}\left[1+\frac{\mu^2 r_h^2}{2}+\frac{m^2}{\nu+1}\left(\frac{r_h}{L}\right)^{\nu+4}\right]\;.
\ee
For $\nu<1$ the energy density can become negative. The temperature of the black brane can be found as usual to be
\be
T=\frac{|f'(r_h)|}{4\pi}=\frac{1}{4\pi r_h}\left[3-m^2\left(\frac{r_h}{L}\right)^{\nu+4}-\frac{\mu^2r_h^2}{2}\right]\;.
\ee
We see that there is an extremal value of the graviton mass $m_*$ at which the temperature of the black brane vanishes:
\be
m_*^2=\left(\frac{L}{r_h}\right)^{\nu+4}\left[3-\frac{\mu^2r_h^2}{2}\right]\;.
\ee

By evaluating the Euclidean on-shell boundary action\footnote{Upon analytic continuation $\tau=it$, $S_E=-iS_L[g_L]=-S_L[g_E]$ and $A_\tau= -iA_t$. We integrate the volume integral as $\int_{r=0}^{r=r_h}dr$. There are contributions to the on-shell action from both integration limits.} we find that there are divergent contributions due to the graviton mass term of the form $S_\text{bdy}\supset m^2(r/L)^{\nu+1}$. Hence, for $\nu+1<0$ additional counterterms that remove the divergences are needed. In order to find the covariant counterterms expressed in terms of the bulk fields the full procedure of holographic renormalization has to be carried out. However, the finite action can be found by simply removing the divergences with counterterms of the form $S^m_\text{count}=m^2\int d^3x\,\sqrt{\gamma}\,\sum_{n=1}^\infty c_n(r/L)^n$, where $\gamma$ is the metric induced on the boundary and $c_n$ are constant coefficients. They can be fixed in such a way that all the divergent terms cancel out. In practice, only several of the counterterms are needed and most of the coefficients $c_n$ are equal to zero.\footnote{Such renormalization procedure for the dRGT massive gravity was carried out in \cite{Blake:2013bqa}. The full covariant renormalization for the mass Lagrangians with $\nu=\{-2,-3\}$ was done in \cite{Andrade:2013gsa,Taylor:2014tka}.} We note that the coefficients $c_n$ depend neither on the temperature nor the chemical potential and, thus, do not affect the thermodynamics. We find the finite Euclidean on-shell boundary action and identify it with the thermodynamic potential
\be\label{free_en}
\Omega=T\left(S_{\text{bdy}}^E+S_{\text{count}}^E\right)=-\frac{VL^2}{2}\left[\frac{1}{r_h^3}+\frac{\mu^2}{2r_h}-\frac{m^2(
\nu+3)}{L^3(\nu+1)}\left(\frac{r_h}{L}\right)^{\nu+1}\right]+\Omega_0(m^2,c_n)\,,
\ee
where $S_{\text{count}}^E=S^m_{\text{count}}+\frac{1}{2}\int d^3x\,\sqrt{\gamma}\,(4/L)$, $V$ is the area of the spatial boundary, and we denote by $\Omega_0$ a constant contribution due to the renormalization procedure. Given the free energy, we find the pressure of the boundary theory as $p=-\Omega/V$. We see then, that the null energy condition is always satisfied in the boundary theory since 
\be
\varepsilon+p=\frac{L^2}{r_h}\left[\frac{2\pi}{r_h}T+\mu^2\right]>0\,.
\ee
We also find that the charge density and entropy density are given by the usual expressions: $q=\mu^2L^2/r_h$ and $s=2\pi L^2/r_h^2$. Together with the energy density given in \eqref{endens} they satisfy the first law of thermodynamics $\varepsilon +p=sT+\mu q$. 

\section{Viscoelastic Properties of the Holographic Solids}
In this section we find the shear viscosity \eqref{viscosity} and elastic modulus \eqref{elasticity} in the field theory dual to the bulk massive gravity with the mass function
\be\label{mass}
M^2(r)=\frac{1}{2L^{2}}\left(\frac{r}{L}\right)^\nu\,.
\ee
By comparing to \eqref{m_ten} and \eqref{mass0} we see that this definition corresponds to the mass potential $V(X)=\frac{1}{n}X^n$ with $n=(4+\nu)/2$.

\subsection{Numerical results}
From the equation of motion \eqref{spin2} it follows that in the near-boundary region the metric perturbations $h\equiv h_\omega$ behave as\footnote{More precisely, the asymptotic UV expansion of the metric perturbations $h$ reads:
\begin{equation*}
\lim_{r\to 0}\,h=h_0\,\left(1+h_p\,\left(\frac{r}{L}\right)^p+\dots\right)+\left(\frac{r}{L}\right)^3\,h_3\,\left(1+\dots\right)+\dots
\end{equation*}
For $\nu<-1$ we have $p<3$ (for example, $p=2$ for $\nu=-2$). However, this does not affect the identification of $h_0$ as the source and $h_3$ as the v.e.v. of the operator associated to the bulk field $h$ and the consequent definition of the Green's function given in \eqref{sublead}.
\color{black}}
\be
\lim_{r\to 0}\,h=h_0+\left(\frac{r}{L}\right)^3h_3+\dots
\ee
showing that the scaling dimension of $h$ is $\Delta = 3$ and is independent on the radial dependence of the graviton mass. The gauge/gravity duality prescription then allows one to find the retarded Green's function as the ratio of the subleading to leading mode of the graviton (see \emph{e.g.} \cite{Hartnoll:2009sz}):
\be\label{sublead}
{\cal G}^R_{T_{ij}\,T_{ij}}=\frac{2\Delta-d}{2L}\,\frac{h_3}{h_0}
\ee
where $d=3$ is the number of spatial dimensions. We numerically solve the equation of motion for the graviton and extract the retarded Green's function by using the above expression. 

\begin{figure}[h!]
\center
\includegraphics[width=6.5cm]{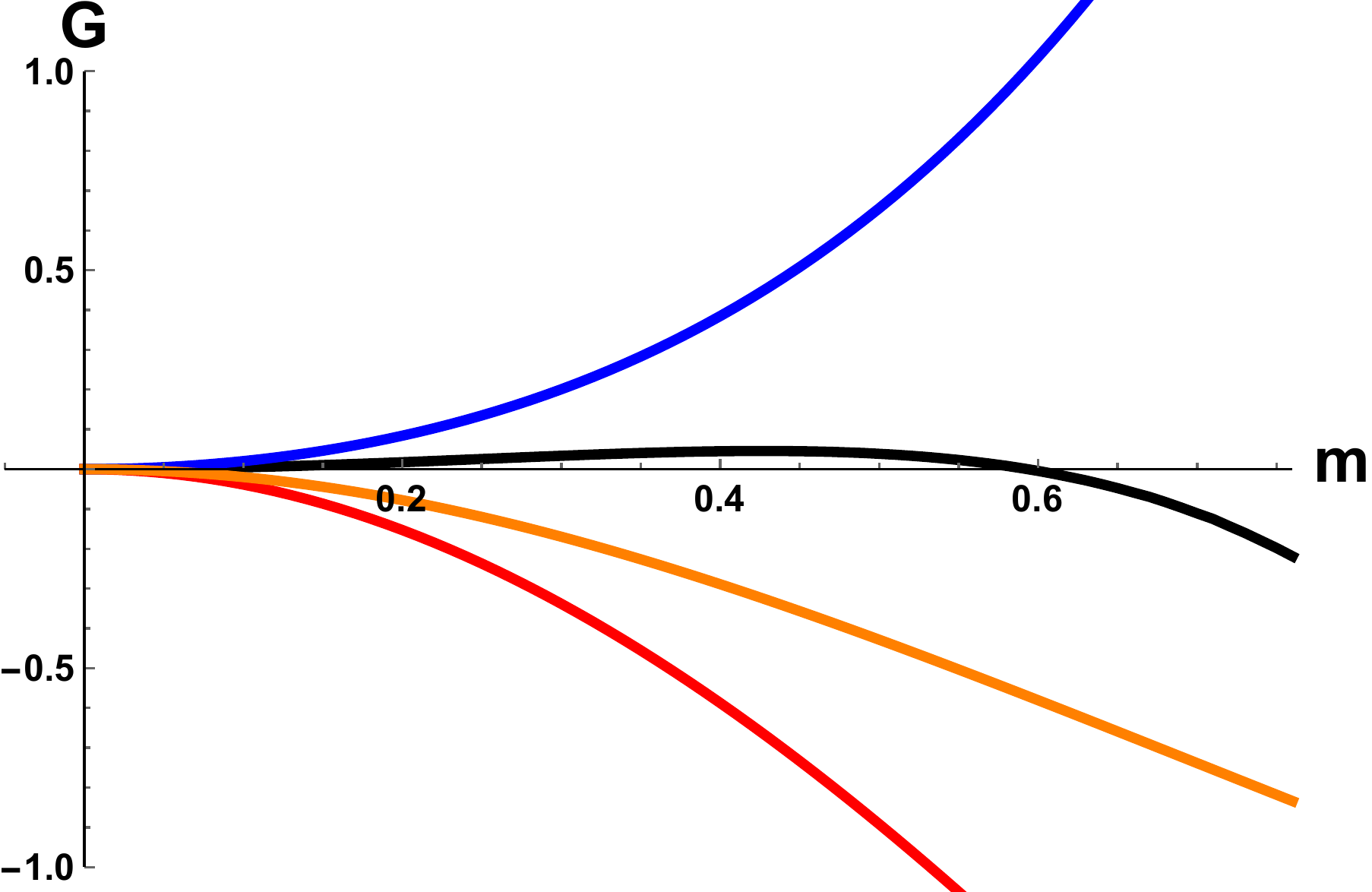}%
\qquad
\includegraphics[width=6.5cm]{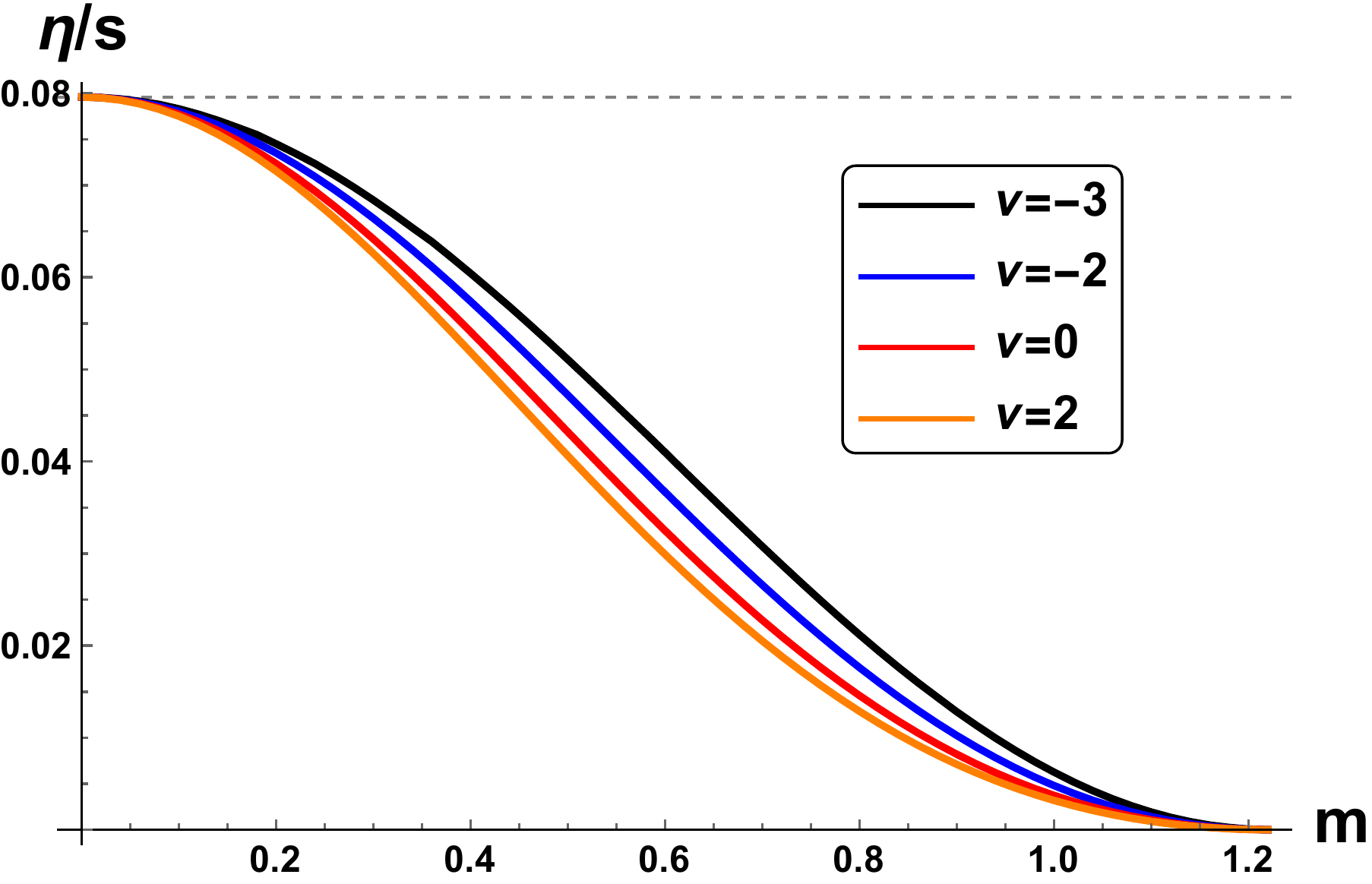}
\caption{Backreacted model $V(X)=X^n$ for $r_h=1$ and zero charge $\mu=0$ for different $n=\frac{4+\nu}{2}$, $\nu=-3,-2,0,2$. \textbf{Left:} Elasticity; \textbf{Right:} $\eta/s$ ratio, the horizontal dashed line shows the value $\eta/s=1/(4\pi)$.}
\label{fig:main}
\end{figure}
In Fig. \ref{fig:main} we show the real part of the Green's function and the viscosity to entropy density ratio as a function of the graviton mass for different values of the exponent~$\nu$. We first observe that the $\eta/s$ ratio goes below the universal value $1/(4\pi)\approx 0.08$ for graviton mass parameter values $m>0$ and thus violates the KSS bound. As expected, in the fluid regime with $m=0$ we recover the standard universal value \eqref{univ}. The second observation that we make is that the real part of the Green's function becomes negative for all values of $\nu$ apart from $\nu=-2$. Although, negative modulus of elasticity can, in principle, be observed in nature it is always associated with instabilities (see \emph{e.g.} \cite{nature}). From the holographic perspective, the fact that there is an instability is not so surprising because the kinetic terms for the St\"uckelberg fields are non-canonical for $V(X)= X^n$ with $n>3/2$ and $n=1/2$ (corresponding to $\nu>-1$ and $\nu=-3$ respectively). Both the numerical and analytical results give a positive rigidity modulus for the canonical St\"uckelberg case,  $n=1$ ($\nu=-2$) with $V=X$, which can therefore be singled out as the most reasonable model from the phenomenological point of view.

We would like to point out that the fact that the KSS bound can be violated in theories with massive gravity duals was also noticed in \cite{Davison:2014lua} for the case $\nu=-2$ corresponding to $V= X/2$ and $\mu=0$. However, this observation was made only at one particular value of the graviton mass $m=\sqrt{2}r_h^{-1}$ for which the energy density \eqref{endens} vanishes\footnote{The expression for the energy density \eqref{endens} is given for the models $V=X^n$. To account for the factor $1/2$ in the model considered in \cite{Davison:2014lua} we need to replace $m^2\to m^2/2$ in \eqref{endens}. It then follows that the energy density vanishes at $m=\sqrt{2}r_h^{-1}$.} and $m/T=\sqrt{8}\pi$. It was then argued by the authors that this result is irrelevant for the physical viscosity due to the fact that for graviton masses of order $m/T\gtrsim 1$ the dual field theory does not admit a coherent hydrodynamic description. Instead a crossover from the coherent hydrodynamic phase of the system to an incoherent regime occurs for graviton mass that is comparable to the black brane temperature. In the results presented in this paper we see the violation of the KSS bound also at arbitrary small values of the graviton mass where the hydrodynamic description applies. We therefore believe that our findings are physically significant and suggest that the KSS bound can be violated in materials with non-zero elastic response. In general, however, we find that the question of whether or not the black branes are close to having a hydrodynamic description is not particularly relevant in the context of holographic solids. In these systems we do not expect the dynamics to be understood in terms of hydrodynamics while there does exist a well defined low energy effective field theory description of solids defined as an expansion at low frequencies and momenta.

\subsection{Analytic estimate}\label{sec:analytics}
In this section we shall provide an analytic derivation for the elasticity and viscosity. 
For this we shall solve the equation of motion \eqref{spin2} for the metric perturbation $h$ in the zero frequency limit by $(i)$ imposing ingoing boundary conditions at the horizon and $(ii)$ requiring that near the AdS boundary the bulk field $\left.h\right|_{r\to 0}=h_0$. The on-shell boundary action can then be brought in the form
\be\label{bdy_split}
S_{\text{bdy}}=S_{r_h}+S_{\epsilon}=\int d^2x\int\frac{d\omega}{2\pi}\,h_0\,\mathcal F(\omega,r)h_0\Big|^{r=r_h}_{r=\epsilon}\;.
\ee
According to the prescription given in \cite{Son:2002sd} the retarded Green's function ${\cal G}^R_{T_{ij}T_{ij}}$ can then be extracted from the on-shell boundary action as 
\be\label{green}
{\cal G}^R_{T_{ij}T_{ij}}=-\lim_{\epsilon\to \,0}2\,\mathcal F(\omega,r)\Big|_{r=\epsilon}\;,
\ee
where the function $\mathcal F(\omega, r)$ is only evaluated at the AdS boundary while the contribution from the horizon is neglected. 

The equation of motion \eqref{spin2} for $h$ in the zero frequency limit can be solved by the following ansatz
\begin{equation}\label{hsol}
h(r)\,=\,h_0\,e^{-\frac{i\,\omega}{4\,\pi\,T}\,\log{f}}\,\left(\Phi_0(r)\,+\,\frac{i\,\omega}{4\,\pi\,T}\,\Phi_1(r)\,+\,\dots\right)\;.
\end{equation}
The ingoing boundary conditions are satisfied by the exponential ansatz while the functions $\Phi_0$ and $\Phi_1$ are required to be regular at the horizon. Near the boundary we demand that 
\begin{equation}
\Phi_0(0)\,=\,1\,,\hspace{1cm}\Phi_1(0)\,=\,0\,.
\end{equation}
The equations for $\Phi_0$ and $\Phi_1$ are obtained by solving the original equation of motion \eqref{spin2} order by order in the frequency $\omega$:
\begin{align}\label{Phi0}
&\frac{z^2}{f}\left(\frac{f}{z^2}\Phi_0'(z)\right)'-\frac{4 \,m^2\, r_h^4\, z^2\, M^2(z)}{L^2f(z)}\,\Phi_0(z)\,=\,0\,,\\\label{Phi1}
&\frac{z^2}{f}\left(\frac{f}{z^2}\Phi_1'(z)\right)'-\frac{4\, m^2\, r_h^4\,z^2\, M^2(z)}{L^2f(z)}\,\Phi_1(z)-\frac{2 \,f'}{f}\Phi_0'(z)+\Phi_0(z) \left(\frac{2 \,f'}{z
   \,f}-\frac{f''}{f}\right)\,=\,0\,,
\end{align}
where we have introduced the variable $z=r/r_h$ and the primes denote derivatives with respect to $z$. 

We note that the last summand in the equation for $\Phi_1$ vanishes on the Schwarzschild-de Sitter background $f(z)=1-z^3$. This simplifies the calculations and corresponds to the probe limit of the full massive gravity theory where we neglect the backreaction of the graviton mass and of the $U(1)$ field on the background metric. We shall perform the calculations in this limit. A  comparison between the numerical results obtained in the proble limit with the corresponding results obtained in the full model including the backreaction is shown in Fig. \ref{fig:backreaction} for $\nu=-2$. As expected, we see that the effect of backreaction is negligible for small values of the graviton mass and introduces no qualitative change to the probe limit results. 

\begin{figure}
\center
\includegraphics[width=6.5cm]{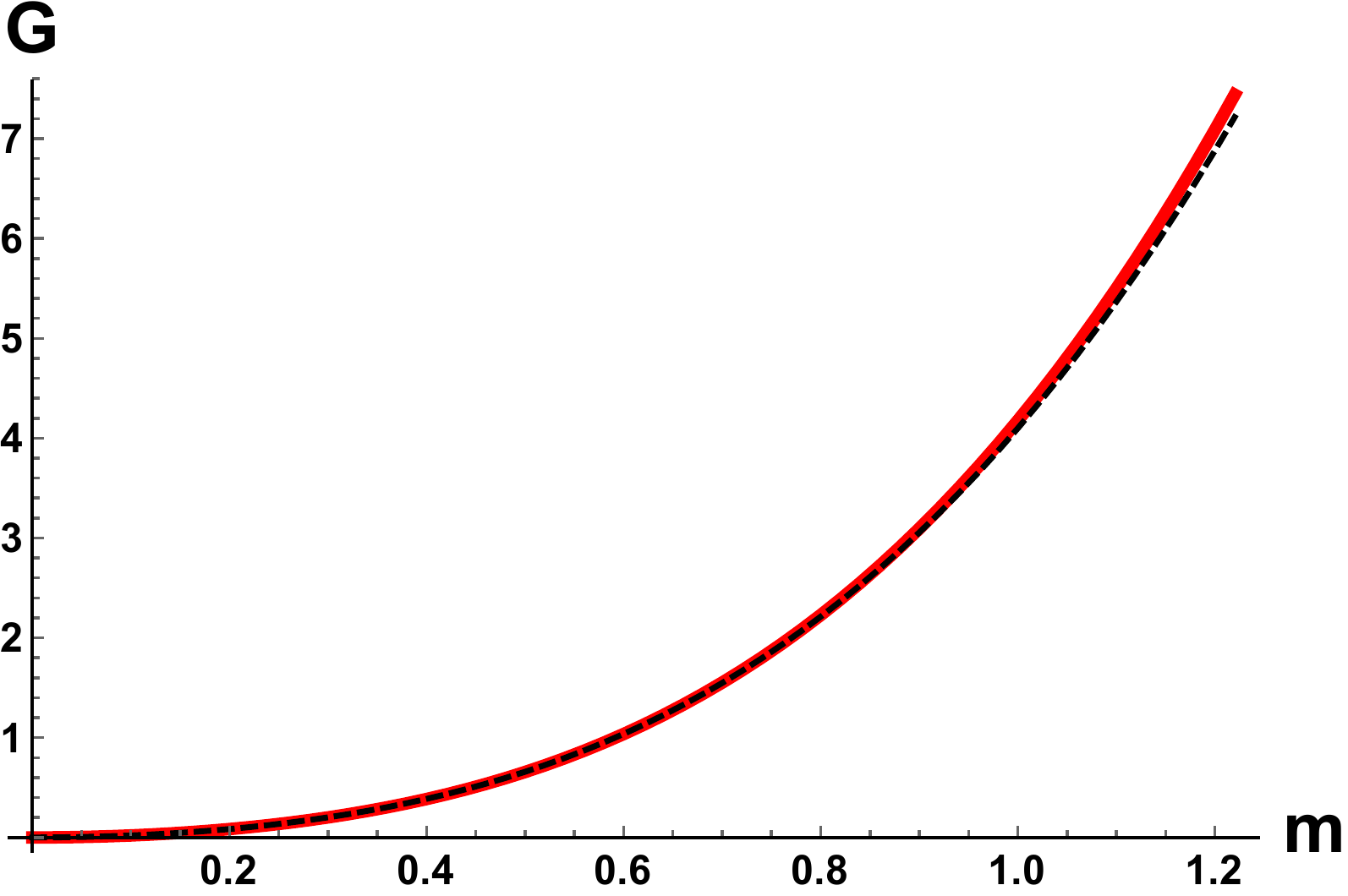}%
\qquad
\includegraphics[width=6.5cm]{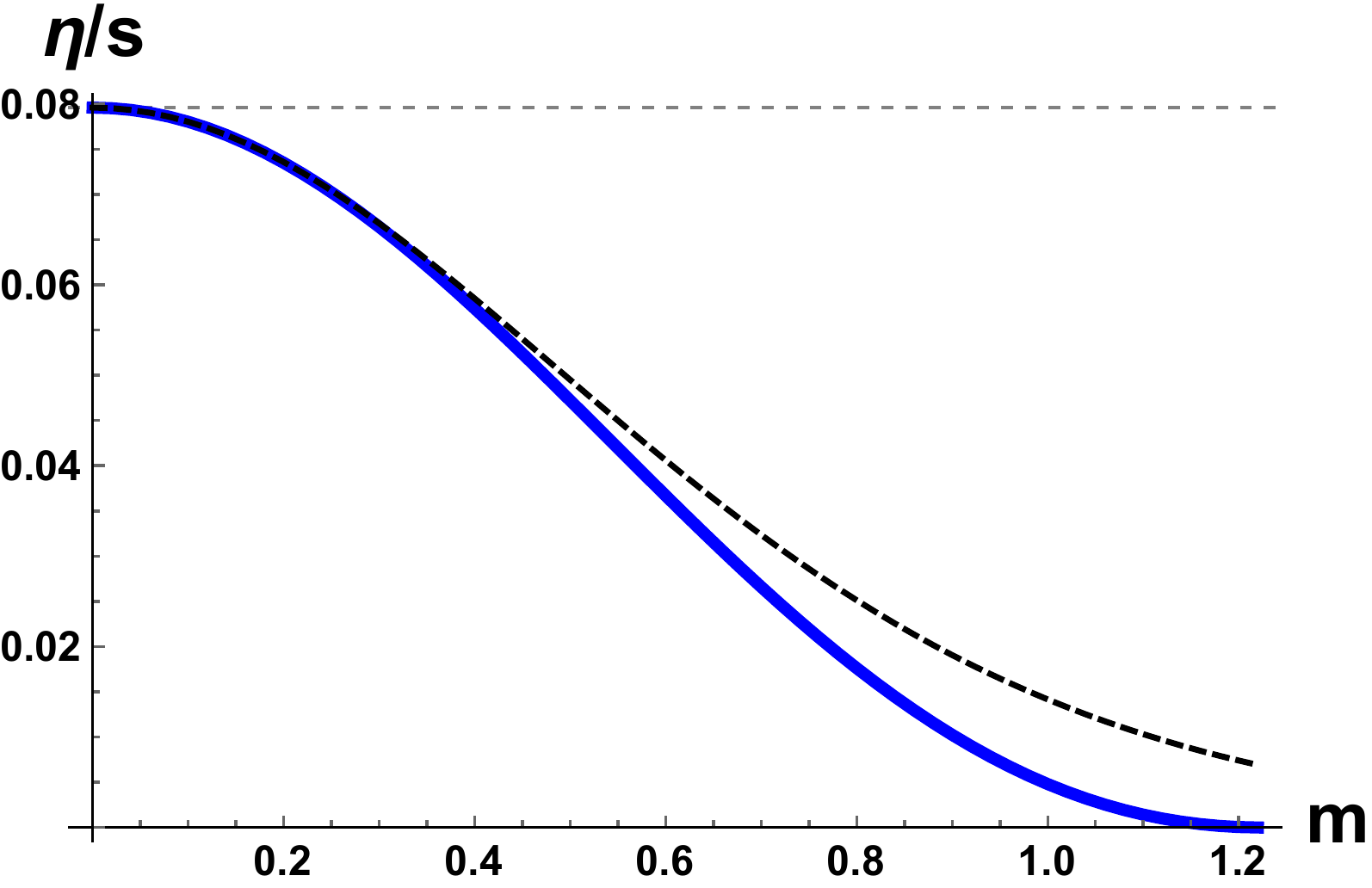}
\caption{Comparison between the numerical results (solid lines) in the backreacted model $V(X)=X$ (with $\nu=-2$) and the probe limit results (dashed lines) for $r_h=1$ and zero chemical potential $\mu=0$. \textbf{Left:} the real part of the Green's function; \textbf{Right:} $\eta/s$ ratio.}
\label{fig:backreaction}
\end{figure}

We consider the mass function \eqref{mass} which now becomes $M^2(z)=z^\nu(r_h/L)^{\nu}/(2L^2)$ 
and use an ansatz for the functions $\Phi_0$ and $\Phi_1$ that is perturbative in the dimensionless graviton mass parameter $m^2$:
\begin{align}\label{Phisol}
\Phi_0=\sum_{n=0}^\infty m^{2n}\phi_n\;,\qquad \Phi_1=\sum_{n=0}^\infty m^{2n}\psi_n
\end{align}
with $\phi_0=1$ and $\psi_0=0$. The equations of motion for $\phi_1$ and $\psi_1$ follow from \eqref{Phi0} and \eqref{Phi1} and read:
\begin{align}\label{eq_phi}
\left(\frac{f}{z^2}\phi_1'\right)'&=4\frac{r_h^4}{L^2}M^2(z)=2\left(\frac{r_h}{L}\right)^{4+\nu}\,z^\nu\,,\\\label{eq_psi}
\left(\frac{f}{z^2}\psi_1'\right)'&=2\frac{f'}{z^2}\phi_1'\,.
\end{align}
We emphasize that the equation \eqref{eq_phi} is exact for any form of the emblackening factor $f(z)$ whereas the equation \eqref{eq_psi} is only valid for the emblackening factor $f(z)=1-z^3$.

The quadratic on-shell boundary action up to real contact terms is given by
\be\label{act_bdy_0}
S_{\text{bdy}}\,=-\,\frac{L^2}{4r_h^3}\int d^3x\,\left(\frac{f}{z^2}\,h(z)\,h'(z)\right)\Bigg|^{z=1}_{z=\epsilon}\;.
\ee
The contribution from the AdS boundary, $z=\epsilon$ with $\epsilon\to 0$, up to first order in the graviton mass squared and frequency reads
\begin{equation*}
S_{\epsilon}=\int d^3x\,\frac{L^2h_0^2}{4\,r_h^3}\left[m^2\left(\frac{f}{\epsilon^2}\phi_1'\right)+\frac{i\omega}{4\pi T}\left(3+m^2\left(\frac{f}{\epsilon^2}\psi_1'\right)-2\,m^2\log f\left(\frac{f}{\epsilon^2}\phi_1'\right)\right)\right]\,.
\end{equation*}
As we shall see below the last term has a non-zero contribution only in the case when $\nu=-4$ which corresponds to the mass potential $V(X)=V_0$ equivalent to a cosmological constant. Since we do not want to modify the cosmological constant and the corresponding radius of the asymtptotically AdS we shall not consider the case $\nu=-4$. 

\subsubsection{The real part: elasticity}
Up to first order in $m^2$ the real part of the Green's function \eqref{green} is determined solely by the derivative of the function $\phi_1$ as:
\be
G=\text{Re} \,{\cal G}=\frac{L^2}{2r_h^3}\,\lim_{\epsilon\to\, 0}\,m^2\left(\frac{f}{\epsilon^2}\phi_1'\right)\;.
\ee
For any given $M^2(z)$, the equation of motion \eqref{eq_phi} can be integrated as
\be\label{eq_phi02}
G= -2 {m^2 r_h}\lim_{\epsilon\to\, 0}  \int_{\epsilon}^{1} M^2(z)\, dz\;.
\ee
For the mass function \eqref{mass} this gives
\be\label{eq_phi2}
\frac{f}{z^2}\phi_1'=c_\nu\left(-z^{\nu+1}+1\right)\,,\qquad\nu\neq -1\,
\ee
with the integration constant $c_\nu$ fixed by demanding that the function $\phi_1(z)$ is regular at the horizon $z=1$ (as assumed implicitly in \eqref{eq_phi02}):
\be\label{cnu}
c_\nu=\frac{-2}{\nu+1}\left(\frac{r_h}{L}\right)^{4+\nu}\,.
\ee
We note that in the near boundary limit $\epsilon\to0$ the expression \eqref{eq_phi2} is divergent for $\nu<-1$. 
This signals the breakdown of the perturbative method, and one has to invoke a proper regularization method in  order to reproduce the finite results that are obtained numerically. There are several ways to do that including the matching asymptotic series approach used in \cite{Policastro:2001yc}, the holographic renormalization along the lines of \cite{Papadimitriou:2004ap,deHaro:2000vlm}, and others. Here we exploit the fact that the expression \eqref{eq_phi2} is finite in the $\epsilon\to0$ limit for $\nu>-1$ and obtain the result for $\nu<-1$ by analytic continuation in $\nu$. In other words, we take the $\epsilon\to0$ limit assuming $\nu>-1$ and then continue analytically the result in~$\nu$. This method is apparently equivalent to a regularization method where the divergent terms are simply discarded. The real part of the Green's function thus becomes
\be\label{real}
G=\frac{L^2}{2\,r_h^3}\,m^2\,c_\nu\,,\qquad\nu\neq -1\,.
\ee
The comparison of the expression \eqref{real} with the numerical results for $\nu=-2$ is given in Fig. \ref{fig:compare}. It shows a good agreement with the numerical results thus further validating our regularization approach. For the sake of completeness we perform the full holographic renormalization of the action \eqref{action} with $\nu=-2$ in the Appendix \ref{sec:renormalization}. As expected we find that upon addition of proper counterterms the divergent contributions to the Green's function cancel. However, the final expression for the real part of the Green's function obtained by the analytic continuation does not exactly coincide with the results obtained by using the covariant holographic renormalization approach --- the two results differ by finite counterterms. 
Since the analytic continuation method agrees well with the numerical results from the prescription \eqref{sublead}, we find that this gives a good confirmation that the $\text{Re} \,{\cal G}$ at $\omega=0$ extracted in this way has an unambiguous physical meaning. In particular, it can be interpreted as the rigidity modulus. We leave for the future a more thorough analysis that shall clarify the most appropriate choice of the counterterms for this purpose.

\subsubsection{The imaginary part: viscosity}
The imaginary part of the Green's function 
\be
\text {Im}\,{\cal G}=\frac{L^2}{2r_h^3}\,\lim_{\epsilon\to0}\,\frac{i\omega}{4\pi T}\left[3+m^2\left(\frac{f}{\epsilon^2}\psi_1'\right)\right]
\ee
can be found by solving equation the \eqref{eq_psi}. Given \eqref{eq_phi2} this leads to
\begin{equation}
\frac{f}{z^2}\psi_1'\,=\,2\,c_\nu\,\int_{1}^z \frac{f'}{f}\,\left(1-x^{\nu+1}\right)\,dx\;
\end{equation}
where we have set the upper boundary of integration so that the right hand side vanishes on the horizon $z=1$ ensuring that the function $\psi_1(z)$ is regular. In order to evaluate the Green's function we are only interested in the value of the above expression at the AdS boundary $z=0$. By using this result we are now able to find the viscosity to entropy density ratio
\begin{equation}
\frac{\eta}{s}\,=\,\frac{1}{4\,\pi}\left(1+\frac{2}{3}\,c_\nu\,m^2\,\int_{1}^0 \frac{f'}{f}\,\left(1-x^{\nu+1}\right)\,dx\right)
\end{equation}
where we have used the relation \eqref{viscosity} and the fact that the temperature of the Schwarzschild-AdS solution with the emblackenig factor $f(z)=1-z^3$ is simply $T=3/(4\pi r_h)$ and the entropy density is given by $s=2\pi(L^2/r_h^2)$. The integral in the above expression can be written in terms of the Harmonic numbers
\be
\mathcal H_{\frac{1}{3}(\nu+1)}\equiv\int_{1}^0 \frac{f'}{f}\,\left(1-x^{\nu+1}\right)\;.
\ee
For the value $\nu=-2$ this gives the exact expression
\be\label{etas-2}
\frac{\eta}{s}\,=\,\frac{1}{4 \pi }-m^2 \,\left(\frac{r_h}{L}\right)^2\, \left(\frac{\log (3)}{\pi }-\frac{1}{3\, \sqrt{3}}\right)\;.
\ee
The quantity in the brackets is positive and hence we see that the viscosity to entropy density ratio is less than the universal value $1/(4\pi)$ thus violating the KSS bound. In Fig.~\ref{fig:compare} we compare the numerical results for the real part of the Green's function and for the $\eta/s$ ratio with the corresponding analytic expressions \eqref{real} and \eqref{etas-2}. We see a good agreement for small values of the graviton mass parameter $m$.


\begin{figure}
\center
\includegraphics[width=6.5cm]{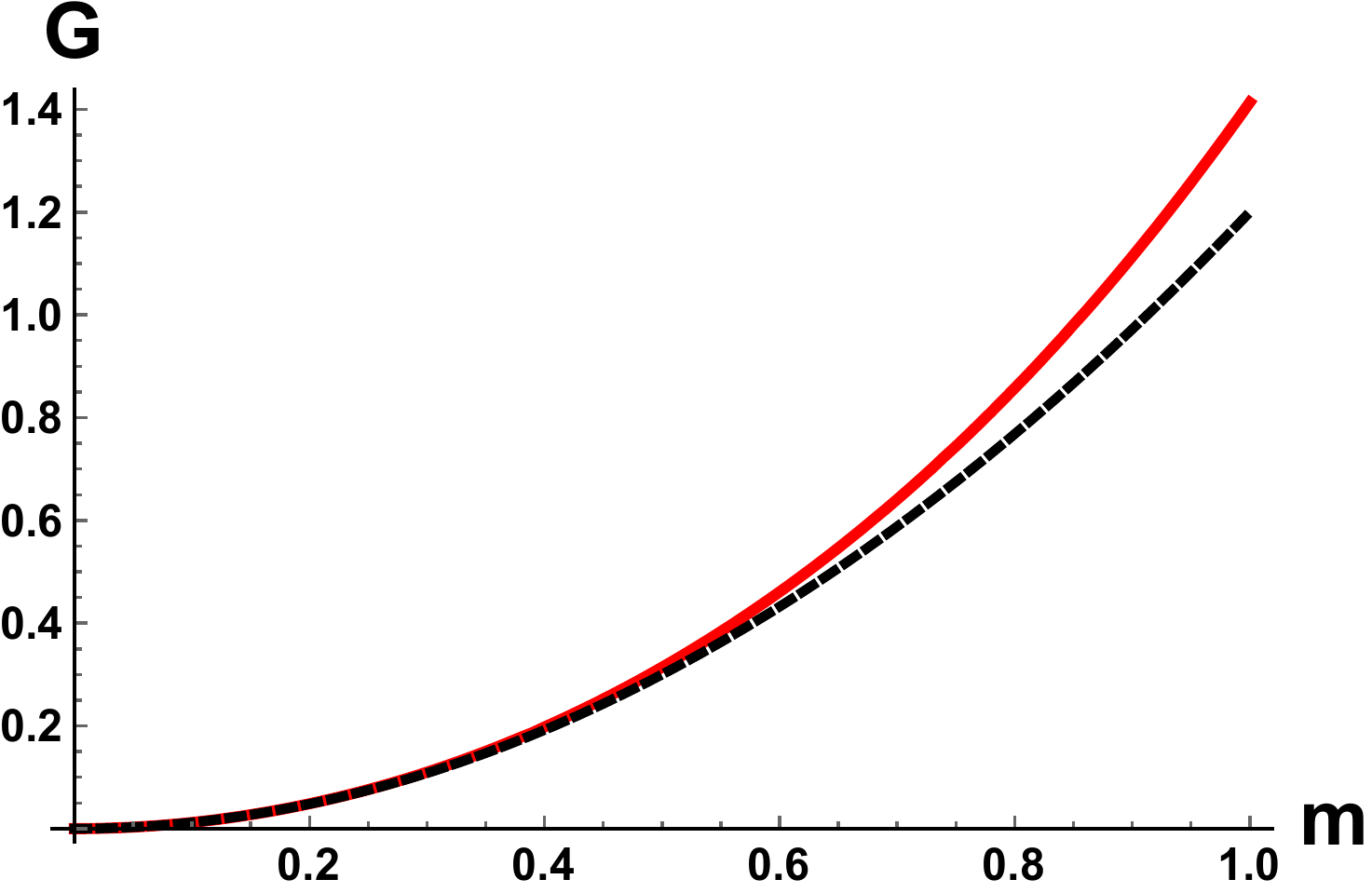}
\qquad
\includegraphics[width=6.5cm]{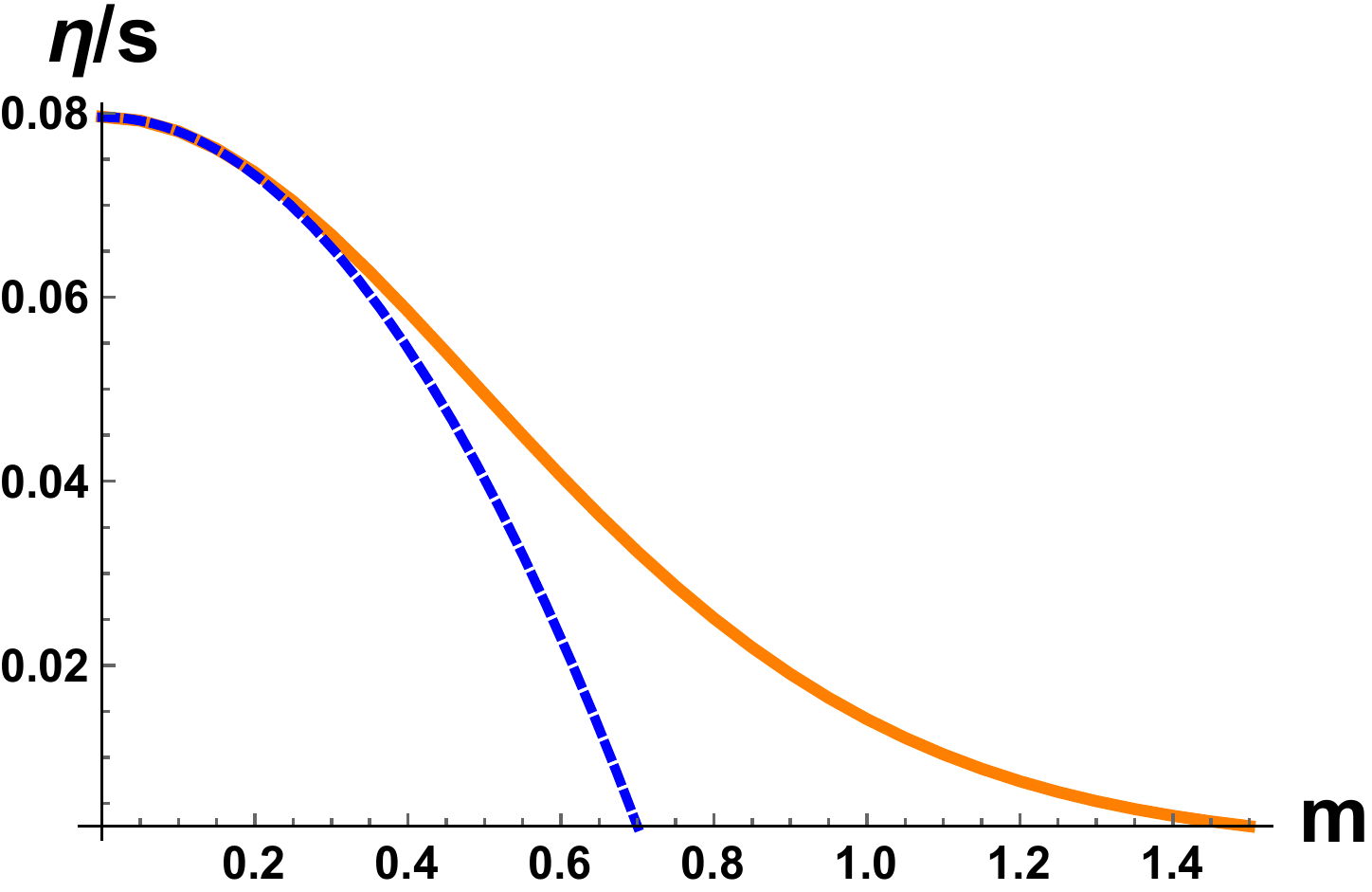}
\caption{Comparison between the numerical results for the Schwarzschild--AdS case and analytic results as a function of the dimensionless graviton mass parameter $m$ for the mass function with $\nu=-2$. Solid and dashed lines are the numerical and analytic results respectively. \textbf{Left:} Real part of the retarded Green's function for $r_h=0.6$. \textbf{Right:} $\eta/s$ ratio for $r_h=1$.}
\label{fig:compare}
\end{figure}

\section{Discussion}
In this work we have shown that in massive gravity theories of the solid type, with the mass potential given by \eqref{vx}, the asymptotically AdS black branes do not obey the KSS lower bound, $\eta/s\geq1/(4\pi)$ \cite{Kovtun:2004de}. Additionally, we have seen a clear correlation between the bound violation and the presence of a non-zero shear elastic modulus. We have given both numerical and analytical evidence that are in complete agreement with each other. We have performed a number of consistency checks to ensure that the result is physical. In the regime where the theory is completely under control (\emph{i.e.} free of instabilities and other  pathologies) the result holds and it is unambiguous. Therefore, we conclude that at least in terms of the Green--Kubo definition of the viscosity \eqref{viscosity} 
there is a physical violation of the KSS bound in the solid massive gravity black branes. There are many potential implications of this result. 

First, it becomes clear that the universal value $\eta/s=1/(4\pi)$ exhibited in Einstein gravity hinges upon a key dynamical property of that theory, 
namely, that the mass of the spin-two graviton mode, as defined in equation \eqref{spin2}, vanishes.
In the gravitational theory, the parameter $\eta$ relates directly to the spin-two absorption cross section \cite{Policastro:2001yc}. Hence, it is not surprising that $\eta$ gets suppressed in the presence of a non-zero mass for the spin-two graviton. Interestingly, the universal result for the viscosity to entropy density ratio holds also in the very large class of the \emph{fluid} type massive gravity theories with mass potential $V=V(Z)$, since they give a zero graviton mass as can be seen from equation \eqref{m_ten}. Thus, in massive gravity theories the $\eta/s$ ratio is very sensitive to the spin-two mass $m^2M^2(r)$: once it is non-zero, the value of the $\eta/s$ ratio is not universal. Instead, it also depends on temperature and other parameters of the theory and interpolates between the KSS value at $m=0$ and zero at large values of $m$.\footnote{In the limit $T/m>>1$, when the temperature is the dominant scale in the system the graviton mass becomes irrelevant and the usual results ($\eta/s=1/4\pi$) are recovered (see \cite{Baggioli:2015gsa} for similar analysis).} We would also like to stress that both the fluid and the solid types of massive gravity theories are dual to theories with momentum dissipation. However, the conjectured KSS bound seems to hold in the case of fluids but is violated in the case of solids. The latter argument therefore shows that there is no direct correlation between the introduction of momentum dissipation (\emph{i.e.} translational symmetry breaking) and the violation of the viscosity to entropy bound.

In the dual field theory picture,  $\eta/s$ is the usual viscosity to entropy ratio of the CFT. From \cite{Alberte:2015isw}, we know that the spin-two graviton mass $m^2M^2(r)$ controls directly the shear elastic modulus or modulus of rigidity, $G$. 
In field theory language, the previous result can then be phrased as follows: once the system exhibits a non-zero modulus of rigidity, $G$, the KSS bound $\eta/s\geq1/4\pi$ is violated. In other words, {\em the KSS bound does not apply to holographic solids}, incarnated as solid massive gravity black branes. 
Let us now discuss whether or not some form of this statement can hold for real solids.

In the first place, one can ask why should $\eta/s$ be allowed to go down once the elasticity $G\neq0$ is switched on? How can a solid present a better fluidity than the most perfect fluids? At this point it is already clear that we are dealing with materials that have both viscosity and elasticity and thus qualify as viscoelastic materials in the sense that we introduced in Section~\ref{sec:viscous}. As emphasized there, the mechanical response for these materials is more complex than for fluids or perfectly elastic solids, and there seems to be a sense in which these materials are capable of flowing  more easily than viscous fluids. 
Some of the particularities of the viscoelastic response are better understood 
by looking at the response of the material under a time-dependent applied stress.
In this regard, some of the properties of viscoelastic materials include the so-called \emph{creep} and \emph{stress relaxation}. Once an applied stress is removed instantaneously, the material relaxes (flows) back to its equilibrium position. Hence, viscoelastic materials, in a way, are able to {\em flow} without any applied stress for some time. This situation formally corresponds to $\eta=0$, so perhaps this is the physical reason why $\eta/s$ can be small. The bottom line is that the more complex viscoelastic response might be the physical explanation for the violation of the KSS bound.
%

In view of this, we shall entertain the possibility that the violation of the $\eta/s\geq1/4\pi$ bound is physical and occurs in more general and realistic materials than just the holographic solids discussed in this paper. We anticipate that there are three types of systems where this can apply:
{\bf i)} strongly correlated solids that admit an effective description in terms of a strongly coupled QFTs with non-zero rigidity; {\bf ii)} materials that are described by weakly coupled QFTs with non-zero rigidity;\footnote{It is however known that at weak coupling the $\eta/s$ ratio considerably exceeds the KSS value in the case of $G=0$ \cite{Kovtun:2004de}. It therefore seems not so easy that the bound is violated in these cases.} 
{\bf iii)} general viscoelastic materials.
While we do not have a convincing evidence for the violation of the KSS bound in the cases~ii) and~iii), the holographic computation gives a strong support that it does indeed occur in the case~i). Interestingly, graphene does comply with the two conditions of this case: it has an enormous rigidity modulus and it contains strong correlations. Our results thus seem to suggest that the KSS bound might be violated in graphene, although it might be that this depends on other factors such as the degree of disorder or on temperature, as we have seen in the holographic computation.  
Earlier theoretical calculations show unexpectedly low values of the $\eta/s$ ratio for graphene, even if satisfying the KSS bound \cite{Mueller:graphene}. However, to the best of our knowledge, no experimental measurement of the $\eta/s$ ratio has been done so far. A recent theoretical proposal for methods of measuring this ratio in solid--state devices was put forward in \cite{Polini} giving a hope for more experimental results in the future.

\begin{figure}[h]
\center
\includegraphics[width=7.5cm]{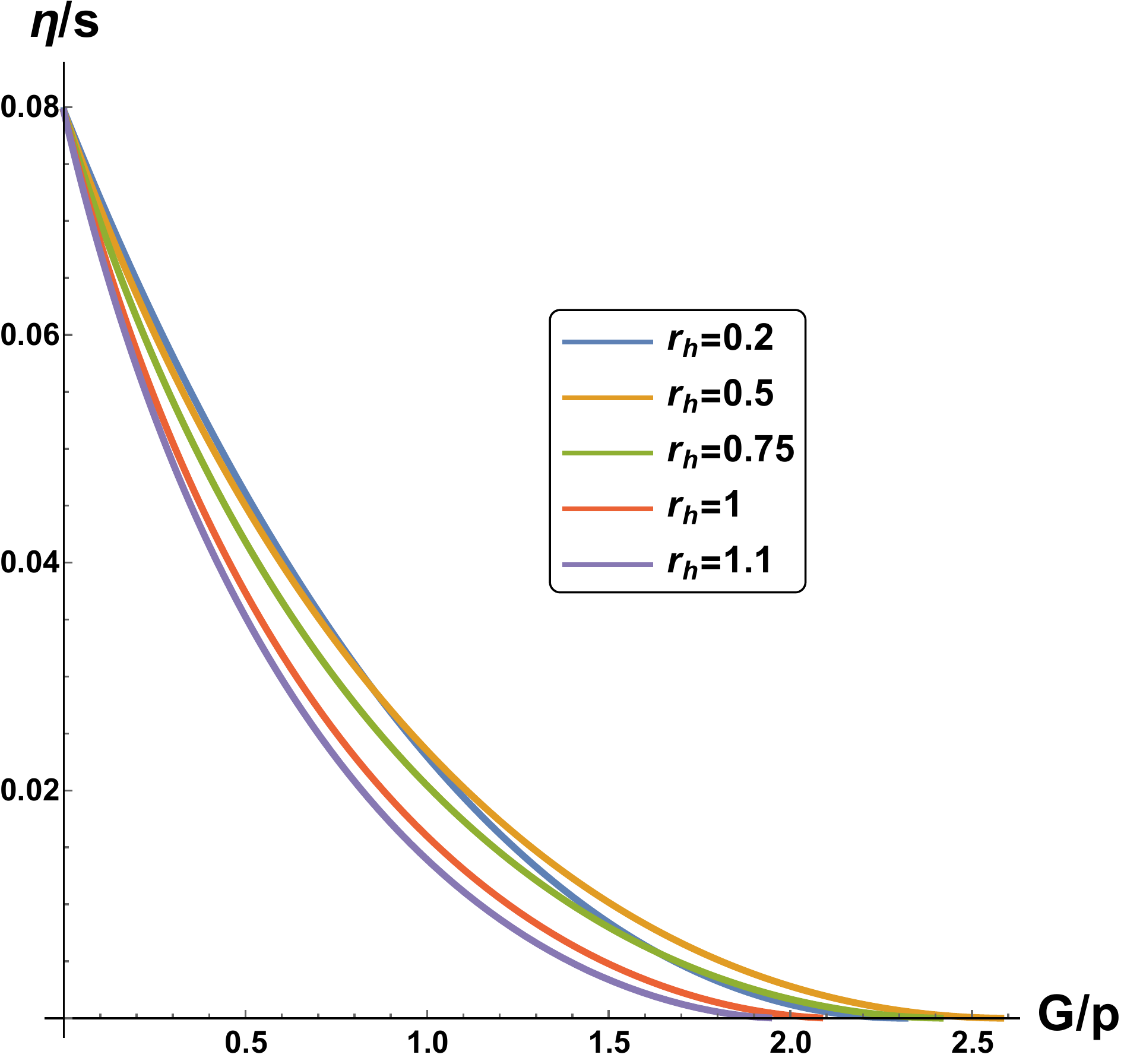}
\caption{Viscosity--elasticity diagram for the $V(X)=X$ backreacted model at fixed chemical potential $\mu=1$ and different values of $r_h$. The value of the graviton mass and the temperature are changing along the solid lines, with $m=0$ at the point where $\eta/s=1/4\pi\approx 0.08$. Similar plots are obtained by keeping $m$ constant and varying $T$ only. }
\label{fig:diagram}
\end{figure}

In the spirit of the KSS conjecture, one can also wonder whether or not there is any generalization of it that holds in solid systems. From dimensional analysis it is reasonable to expect that if there does exist a more general bound, it should involve the rigidity to pressure ratio, $G/p$, in addition to the $\eta /s$ ratio. 
%
In Fig.~\ref{fig:diagram} we plot $\eta/s$ against $G/p$ for the holographic solid with $\nu=-2$ and see a clear correlation. 
Keeping the KSS logic \cite{Kovtun:2004de} that the gravity solutions might represent the least dissipative materials, the Fig.~\ref{fig:diagram} then suggests that there might be a more general bound in (viscoelastic) solids. At relatively large temperatures 
this would approximately take the form   
$$4\pi\frac{\eta}{s} + \mathcal{C}  \frac{G}{p} \geq 1$$
with  $\mathcal{C}$ being an order-one constant. 
We further note that at zero temperature, $T=0$, the viscosity to entropy ratio becomes zero and the rigidity modulus reaches its maximum value. The existence of a similar universal bound on the thermoelectric transport coefficients has been recently conjectured in \cite{Kovtun:2014nsa,Hartnoll:2014lpa} and explored in holographic theories featuring momentum dissipation in \cite{Amoretti:2014ola}.

%

\section*{\sc Note added}

While this work was being completed Ref.~\cite{Hartnoll:2016tri} appeared, where the violation of the KSS bound in massive gravity theories is also discussed.Near the final stages of our work we learned about similar findings of yet another collaboration that appeared later in \cite{Burikham:2016roo} after our work was already made public. Where our results overlap, they agree.

\section*{Acknowledgments}
We thank Andrea Amoretti, Danny Brattan, Mikhail Goykhman, Andrei Khmelnitsky, Rene Meyer, Daniele Musso, Nick Poovuttikul, Stephan Roche and Dam Thanh Son for very useful discussions and comments about this work. MB would like to thank University of Illinois, ICMT and Philip Phillips for the warm hospitality during the completion of this work.
We acknowledge support by the Spanish Ministry MEC under grant FPA2014-55613-P and the Severo Ochoa excellence program of MINECO (grant SO-2012-0234), as well as by the Generalitat de Catalunya under grant 2014-SGR-1450.

\appendix
\section{Holographic renormalization for $\nu=-2$}\label{sec:renormalization}
In this section we shall compute the retarded Green's function following the prescription of \cite{Son:2002sd,Policastro:2002se} from the \emph{renormalized} on-shell boundary action. The main motivation for performing the full holographic renormalization procedure is the fact that the on-shell boundary action \eqref{act_bdy_0} that follows from the action \eqref{action} is divergent in the near boundary limit $r\to0$. In section \ref{sec:analytics} we have simply dropped the divergent terms and assumed that the finite results that we derive for the retarded Green's function will not be affected by the renormalization procedure. Here we check this assumption explicitly.

We consider the massive gravity action \eqref{action} with $n=1$ (or equivalently, $\nu=-2$). According to \eqref{mass0} this corresponds to the mass parameter 
\be\label{massA}
M^2(r)=\frac{1}{2L^2}\left(\frac{r}{L}\right)^{-2}\;.
\ee
In this case the divergent on-shell boundary action that follows from \eqref{action} should be supplemented by the counterterm action \cite{Papadimitriou:2004ap,deHaro:2000vlm}: 
\be\label{count}
S_{\text{count}}=S_{\text{count}}^{\text{gr}}+S_{\text{count}}^{\text{m}}=\frac{1}{2}\int_{r\to 0} d^3x\,\sqrt{-\gamma}\left(-\frac{4}{L}-LR(\gamma)+\frac{m^2}{L}\pt_a\phi^I\pt^a\phi^J\delta_{IJ}\right)
\ee
where $a=\{t,x,y\}$ is an index in the boundary theory. This consists of counterterms due to the divergences arising in the Einstein action and in the scalar fields action (the Maxwell action is finite and does not require any additional counterterms).

Evaluating \eqref{action} together with \eqref{count} we find the boundary action\footnote{According to the prescription of \cite{Son:2002sd}, we only consider the near boundary contribution $S_\epsilon$ to the total boundary as in the split \eqref{bdy_split}.}
\begin{align}\label{bdy_act_1}
S_{\text{bdy}}^{\text{tot}}=-\frac{VL^2}{2T}&\left[\frac{1}{r_h^3}+\frac{\mu^2}{2r_h}-\frac{m^2}{r_hL^2}+\frac{1}{2}\left(\frac{1}{r_h^3}+\frac{\mu^2}{2r_h}-\frac{m^2}{r_hL^2}\right)h^2(\epsilon)-\right.\notag\\
-&\left.\frac{m^2}{\epsilon r_hL^2}h^2(\epsilon)+\frac{\omega^2}{2r_h\epsilon}h^2(\epsilon)\right]+\frac{L^2}{4r_h^3}\frac{V}{T}\frac{f(\epsilon)}{\epsilon^2}h(\epsilon)h'(\epsilon)\;
\end{align}
where $\epsilon=r/r_h$. The part quadratic in metric perturbations in the above action differs from the boundary action \eqref{act_bdy_0} used in section \ref{sec:analytics} by finite contact terms and divergent counterterms. The last term in \eqref{bdy_act_1} coincides with the near boundary contribution of \eqref{act_bdy_0}. We note that the background value of the above action 
\be\label{free_en2}
S_0=S_{\text{bdy}}^{\text{tot}}\Big|_{h=0}=-\frac{VL^2}{2T}\left(\frac{1}{r_h^3}+\frac{\mu^2}{2r_h}-\frac{m^2}{r_hL^2}\right)=-\frac{V}{T}\left(p-\frac{m^2}{r_h}\right)\neq\Omega/T
\ee
does not coincide with the value found from the Euclidean action given in \eqref{free_en} pointing towards a mismatch between the Euclidean and Lorentzian prescriptions in the case of massive gravity. The difference only appears in the mass term contribution suggesting that an additional finite counterterm should be added to the mass sector of the counterterm action \eqref{count}. The only choice with not more than two derivatives and preserving the homogeneity and isotropy in the boundary theory that is finite in the limit $\epsilon\to 0$ is
\be\label{finite}
S_{\text{count}}^{\text{finite}}=\frac{\alpha}{2\sqrt{2}}\frac{m^2}{r_h}\int d^3x\,\sqrt{-\gamma}\left(\pt_a\phi^I\pt^a\phi_I\right)^{3/2}=\alpha\frac{V}{T}\frac{m^2}{r_h}\sqrt{f(\epsilon)}\left(1+h^2(\epsilon)\right)\;,
\ee
where the parameter $\alpha$ shall be determined from the background value $S_0$ of the total boundary action. We note that this counterterm involves an explicit dependence on $r_h$ and thus necessarily will affect the thermodynamics of the system. In fact, this is precisely our goal here since the expression \eqref{free_en2} contradicts the Euclidean results. It is not clear whether an addition of such a counterterm is viable, but we shall nevertheless explore this possibility here. The results without the addition of this ambiguous counterterm can be readily recovered by setting $\alpha=0$ in the final expressions.

By combining \eqref{finite} with the boundary action \eqref{bdy_act_1} we obtain
\begin{align}\label{bdy_act_2}
S_{\text{bdy}}^{\text{tot}}=-\frac{VL^2}{2T}&\left[\frac{1}{r_h^3}+\frac{\mu^2}{2r_h}-\frac{m^2(2\alpha+1))}{r_hL^2}+\frac{1}{2}\left(\frac{1}{r_h^3}+\frac{\mu^2}{2r_h}-\frac{m^2(4\alpha+1)}{r_hL^2}\right)h^2(\epsilon)-\right.\notag\\
-&\left.\frac{m^2}{\epsilon r_hL^2}h^2(\epsilon)+\frac{\omega^2}{2r_h\epsilon}h^2(\epsilon)\right]+\frac{L^2}{4r_h^3}\frac{V}{T}\frac{f(\epsilon)}{\epsilon^2}h(\epsilon)h'(\epsilon)\;.
\end{align}
By setting $\alpha=-1$ we recover $S_0=\Omega/T=-pV/T$. 

 
The renormalized action \eqref{bdy_act_2} when evaluated on the solution \eqref{hsol}, \eqref{Phisol}, is expected to be finite in the $\epsilon\to 0$ limit. We check it explicitly for the real part of the action since only the solution \eqref{eq_phi2} is valid also for the full emblackening factor \eqref{fsol} considered here. Upon substituting the solution in the boundary action we obtain:
\be
\text{Re}\,S_{\text{bdy}}^{\text{tot}}=S_0-\frac{V}{2T}\left(p-\frac{m^2(2\alpha+1)}{r_h}\right)h_0^2+\frac{V}{2T}\frac{m^2}{r_h}h_0^2+\mathcal O(m^4,\omega^2)
\ee
The real part of the Green's function \eqref{green} is then $\text{Re}\,\mathcal G=2T\text{Re} (S_{\text{bdy}}^{\text{tot}}-S_0)/h_0^2/V$ which reads
\be\label{realG}
\text{Re }\mathcal G=-\left(p-\frac{m^2(2\alpha+1)}{r_h}\right)+\frac{m^2}{r_h}\;.
\ee
Only the last term is captured by the minimal renormalization procedure carried out in section \ref{sec:analytics}. Indeed, by comparing the last term in the above expression with \eqref{real} with $\nu=-2$ we see that they coincide.
The term in the brackets in \eqref{realG} arises from finite contact terms and in the case of massless gravity reduces to pure pressure. This is a charecteristic feature for fluids \cite{Baier:2007ix}. The second term is what we have defined as elasticity in section \ref{sec:analytics}. We note, however, that upon the choice $\alpha=-1$ the real part of Green's function reduces to $\text{Re }\mathcal G=-p$ as in the case of fluids in \cite{Baier:2007ix}. We thus see that there is some ambiguity in the definition of the real part of the Green's function (it has been pointed out previously also in \cite{Gubser:2008sz}). Nevertheless, the definition of elasticity as proposed in section \ref{sec:analytics} is unambiguous and shows a clear distinction between the cases of holographic solids and massless gravity theories (including holographic fluids).

\end{document}